\documentclass[aps,pra,floatfix,twocolumn,superscriptaddress]{revtex4}
\usepackage{amsmath,amssymb}
\usepackage{graphicx}
\usepackage{epsfig}
\usepackage{psfrag}
\usepackage[usenames]{color}
\usepackage{xcolor}
\usepackage{hyperref}
\usepackage{ulem}

\begin{document}

\title{Coherent photogalvanic effect in fluctuating superconductors}

\author{V.~M.~Kovalev}
\affiliation{A.V.~Rzhanov Institute of Semiconductor Physics, Siberian Branch of Russian Academy of Sciences, Novosibirsk 630090, Russia}
\affiliation{Novosibirsk State Technical University, Novosibirsk 630073, Russia}

\author{K.~Sonowal}
\affiliation{Center for Theoretical Physics of Complex Systems, Institute for Basic Science (IBS), Daejeon 34126, Korea}
\affiliation{Basic Science Program, Korea University of Science and Technology (UST), Daejeon 34113, Korea}

\author{I.~G.~Savenko}
\affiliation{Center for Theoretical Physics of Complex Systems, Institute for Basic Science (IBS), Daejeon 34126, Korea}
\affiliation{Basic Science Program, Korea University of Science and Technology (UST), Daejeon 34113, Korea}

\date{\today}

\begin{abstract}
We develop a theory of the coherent photogalvanic effect (CPGE) in low-dimensional  superconductors in the fluctuating regime.
It manifests itself in the appearance of a stationary electric current of Cooper pairs under the action of two coherent electromagnetic fields of light with frequencies lying in the sub-terahertz range.
We derive the general formula for the electric current density, study the particular cases of linear and circular polarizations of the external light fields, and
show that the current might have a non-monotonous spectrum at certain polarization angles and turns out very sensitive to the proximity of the ambient temperature to the critical temperature of superconducting transition: Approaching the critical temperature, the peak in the spectrum becomes narrower, its frequency experiences a redshift, and the intensity of the peak drastically grows.
\end{abstract}

\maketitle


In a broad sense, the photogalvanic (or photovoltaic) effect (PGE) consists in the emergence of an electric current or voltage in the sample under the action of electromagnetic (EM) field of light. 
Recent years demonstrate a growing interest to PGE, which has been studied in graphene~\cite{RefPEgraphene}, Weyl semimetals~\cite{RefPEWeyl1, RefPEWeyl2}, mono- and dichalcogenides~\cite{RefPEmono, RefPEour},  ferroelectrics~\cite{RefPEFerr1, RefPEFerr2}, and at terahertz frequencies~\cite{RefPETHz}.
There exist various types of PGE, including the bulk PGE~\cite{RefBulk1, RefBulk2}, valve effect in semiconducting p-n junctions, photo–Dember effect~\cite{RefDember}, and photo-piezoelectricity, among others. 
All these phenomena are due to the presence of inhomogeneities either in the media (like a p-n junction) or in the light field itself.

There also exists PGE, which requires neither the inhomogeneity of optical excitation
of electron-hole pairs nor the inhomogeneity of the sample~\cite{Belinicher, IvchenkoPikus, RefBulk1, GanichevPrettl}. 
Instead, the media lacking the inversion center is required.
In this case, the stationary electric current represents the second-order response of the system to a uniform electric field $\mathbf{E}$ with frequency $\omega$, $j_\eta=T_{\eta\lambda\gamma}E_{\omega\lambda}E^*_{\omega\gamma}$, where $\eta,~\lambda,~\gamma=x,~y,~z$. 
Thus, it is determined by the 3-rank conductivity tensor, $T_{\eta\lambda\gamma}$ (which is nontrivial only in the media lacking the inversion center). 
The microscopic origin of this PGE is in the asymmetry of the interaction potential and the electron scattering processes or the crystal-induced Bloch wave function. 
Recently, other mechanisms of PGE related to the trigonal warping of valleys in transition metal dichalcogenides have also been suggested~\cite{Kovalev1, Kovalev2}.

One of the kinds of PGE is the coherent photogalvanic effect (CPGE).
It was predicted in works~\cite{RefEntin, RefBaskinEntin} and observed experimentally in~\cite{RefBalakirev1, RefBalakirev2}. 
It represents the emergence of a stationary electric current in a spatially homogeneous conducting sample exposed to two EM fields with frequencies $\omega$ and $2\omega$.
Phenomenologically, CPGE current can be written as~\cite{RefEntin} $j_\eta=\chi_{\eta\lambda\gamma\delta}E_{2\omega \lambda} E_{-\omega\gamma} E_{-\omega\delta} +\mathrm{c.c.}$, where  $\chi_{\eta\lambda\gamma\delta}$ is a 4-rank generalized conductivity tensor, and $\mathbf{E}=2\mathrm{Re}(\mathbf{E}_\omega e^{i\omega t}+\mathbf{E}_{2\omega} e^{2i\omega t})$, where $\mathbf{E}_{-\omega}=\mathbf{E}_\omega^*$.
As compared with the traditional PGE, CPGE does not require the absence of the inversion symmetry and it depends on the phase difference between the EM fields.

Recently, photoinduced nonlinear transport phenomena in superconductors attract growing interest of the community~\cite{OurKus2}. 
In particular, there has been proposed the photon drag effect~\cite{Boev} and the third harmonic generation~\cite{Silaev}. 
PGEs have not been so far addressed in superconductors, to the best of our knowledge.
It is known that the PGE in normal conductors at low temperatures exists due to the asymmetrical impurity scattering processes of electrons, like the skew- and side-jump effects, among other. 
These processes have been investigated in view of recent research on the anomalous Hall effect in fluctuating superconductors~\cite{Levchenko}, and it has been demonstrated that the Aslamazov-Larkin correction is not dressed by these asymmetric impurity scatterings.

In this paper, we study CPGE in a superconductor in the fluctuating regime~\cite{AL, LarkinVarlamov2005}, when the temperature is slightly above the critical temperature of superconducting (SC) transition $T_c$ and in addition to normal (unpaired) electrons there start to emerge and collapse Cooper pairs called in this case the \textit{SC fluctuations} (SFs) since the density of Cooper pairs fluctuates, according to the Aslamazov-Larkin (AL) effect.
These SFs can dramatically change the conductivity of the system due to an additional \textit{paraconductivity} term.
As we have shown in previous works~\cite{RefAl1}, the presence of SFs can also drastically change the optical response of the system.


To describe the CPGE of SFs, we will use the Boltzmann transport equations approach~\cite{LarkinVarlamov2005}, in the framework of which the Cooper pairs are described by the distribution function and an effective energy-dependent lifetime.
This approach has been proved sufficient if one considers the AL corrections to the conductivity~\cite{Mishonov1, Mishonov2, Mishonov3, Mishonov4}, which we do in this paper.

\pagebreak

\textit{Theory. }
The Boltzmann equation for SFs in the uniform external electromagnetic field reads
\begin{gather}\label{Eqq1}
\frac{\partial f}{\partial t}+2e[\textbf{E}_\omega(t)+\textbf{E}_{2\omega}(t)]\cdot\frac{\partial f}{\partial \textbf{p}}+\frac{f-f_0}{\tau_\textbf{p}}=0,
\end{gather}
where $f$ is the distribution function of fluctuating Cooper pairs, $t$ is time, $e$ is electron charge, $\textbf{E}_\omega(t)=\textbf{E}_\omega e^{i\omega t}+\textbf{E}^*_\omega e^{-i\omega t}$ and $\textbf{E}_{2\omega}(t)=\textbf{E}_{2\omega} e^{2i\omega t}+\textbf{E}^*_{2\omega} e^{-2i\omega t}$ are the first and second harmonics of the electromagnetic field of frequency $\omega$,
$\mathbf{p}$ is the center-of-mass momentum of the Cooper pair with the absolute value $p=|\mathbf{p}|$;
$\tau_{\textbf{p}}=\pi\alpha/(16\varepsilon_\mathbf{p})$ is the effective Cooper pair lifetime with $\alpha$ the parameter of the AL theory~\cite{LarkinVarlamov2005}; $\alpha$ is calculated using the relation $4m\alpha T_c\xi^2/\hbar^2 =1$, where $\xi$ is the correlation length given by
\begin{equation}
    \xi^2 = \frac{v_F^2\tau^2}{2}\Bigg[\psi\Big(\frac{1}{2} \Big)-\psi\Big(\frac{1}{2}+\frac{\hbar}{4\pi T\tau}\Big) + \frac{\hbar \psi'(1/2)}{4\pi T \tau}\Bigg].
\end{equation}
Here $\psi(x)$ is the digamma function; $v_F= \hbar\sqrt{4\pi n}/m$ is the Fermi velocity; $\tau$ is the electron relaxation time,
$\varepsilon_\mathbf{p}=\varepsilon_p=p^2/4m+\mu$ is Cooper pair energy with $m$ the  electron mass, $\mu=\alpha T_c\epsilon$, $\epsilon=(T-T_c)/T_c>0$ the reduced temperature~\cite{LarkinVarlamov2005};
$f_0=T/\varepsilon_p$ is the classical Rayleigh-Jeans distribution of Cooper pairs at temperature
$T$ in the absence of external perturbations.

Furthermore, we assume that the EM fields cause small perturbations of the local density of SFs and do the expansion~\cite{Kittel, Abrikosov}, $f(t)=f_0+\sum\limits_nf^{(n)}(t)$.
The $n-$th order correction obeys the equation
\begin{gather}\label{Eqq2}
\left(\frac{\partial }{\partial t}+\frac{1}{\tau_\textbf{p}}\right)f^{(n)}(t)=-2e[\textbf{E}_\omega(t)+\textbf{E}_{2\omega}(t)]
\cdot
\frac{\partial f^{(n-1)}(t)}{\partial \textbf{p}}.
\end{gather}

The general formula for the CPGE current density is $j_\eta=2e\int d\textbf{p}u_\eta f/(2\pi\hbar)^{2}$, where $\eta=x,~y$ and $u_\eta=p_\eta/2m$ is a Cooper pair velocity.
In our case, the lowest-order nonzero contribution to the stationary current reads
\begin{gather}\label{Eqq3}
j_\eta=2e\int\frac{d\textbf{p}}{(2\pi)^2}u_\eta \langle f^{(3)}(t)\rangle,\\\nonumber
\langle f^{(3)}(t)\rangle=-2e\tau_\textbf{p}\textmd{Re}\,\left(\textbf{E}^*_\omega
\cdot
\frac{\partial f^{(2)}_\omega}{\partial \textbf{p}}+
\textbf{E}^*_{2\omega}
\cdot
\frac{\partial f^{(2)}_{2\omega}}{\partial \textbf{p}}\right),
\end{gather}
where $\langle...\rangle$ stands for the time-averaging and the second-order corrections satisfy
\begin{gather}\label{Eqq4}
\nonumber
\left(i\omega+\frac{1}{\tau_\textbf{p}}\right)f^{(2)}_\omega=-2e\left(\textbf{E}_{2\omega}
\cdot
\frac{\partial f^{*(1)}_{\omega}}{\partial \textbf{p}}+
\textbf{E}^*_{\omega}
\cdot
\frac{\partial f^{(1)}_{2\omega}}{\partial \textbf{p}}\right),\\
\left(2i\omega+\frac{1}{\tau_\textbf{p}}\right)f^{(2)}_{2\omega}=-2e\textbf{E}_{\omega}
\cdot
\frac{\partial f^{(1)}_{\omega}}{\partial \textbf{p}},
\end{gather}
and the first-order corrections read
\begin{gather}\label{Eqq5}
\left(i\omega+\frac{1}{\tau_\textbf{p}}\right)f^{(1)}_\omega=-2e\textbf{E}_{\omega}
\cdot
\frac{\partial f_{0}}{\partial \textbf{p}}=-2e(\textbf{u}\cdot\textbf{E}_{\omega})f_0',\\
\nonumber
\left(-i\omega+\frac{1}{\tau_\textbf{p}}\right)f^{*(1)}_\omega=-2e\textbf{E}^*_{\omega}
\cdot
\frac{\partial f_{0}}{\partial \textbf{p}}=-2e(\textbf{u}\cdot\textbf{E}^*_{\omega})f_0',\\
\nonumber
\left(2i\omega+\frac{1}{\tau_\textbf{p}}\right)f^{(1)}_{2\omega}=-2e\textbf{E}_{2\omega}
\cdot
\frac{\partial f_{0}}{\partial \textbf{p}}=-2e(\textbf{u}\cdot\textbf{E}_{2\omega})f_0',
\end{gather}
where $f'_0=\partial f_0/\partial\varepsilon_p$.
Integrating by parts in Eq.~\eqref{Eqq3} and taking the integrals, introducing for convenience dimensionless variables,
$\kappa=\varepsilon_p/\mu$ and $\beta=\pi\omega/(16T_c\epsilon)$, and then using $\textbf{E}_{\omega}=E_1\textbf{e}_{\omega}$ and $\textbf{E}_{2\omega}=E_2\textbf{e}_{2\omega}$, where we introduce two unity vectors in the directions of electric field harmonics, and finally parameterizing the fluctuating Cooper pair velocity as $\textbf{u}=u\textbf{n}$, where $\textbf{n}=(\cos\varphi,\sin\varphi)$, we can find the total current density (see Supplemental Material~\cite{SM}),
\begin{eqnarray}\label{Eqq14}
j_\eta&=&\chi_{\eta\lambda\gamma\delta}e^*_{\omega\lambda}e^*_{\omega\gamma}e_{2\omega\delta}+
\chi^*_{\eta\lambda\gamma\delta}e_{\omega\lambda}e_{\omega\gamma}e^*_{2\omega\delta}\\
\nonumber
&&+\zeta_{\eta\lambda\gamma\delta}
e^*_{2\omega\lambda}e_{\omega\gamma}e_{\omega\delta}+\zeta^*_{\eta\lambda\gamma\delta}
e_{2\omega\lambda}e^*_{\omega\gamma}e^*_{\omega\delta},
\end{eqnarray}
where
\begin{eqnarray}
\label{EqTensors}
\chi_{\eta\lambda\gamma\delta}&=&\frac{j_0}{2}\int\limits_1^\infty\frac{(\kappa-1)d\kappa}{\kappa^2}\left(\frac{1}{\kappa-i\beta}+\frac{1}{\kappa+2i\beta}\right)\\
\nonumber
&&\times\frac{\partial}{\partial \kappa}\left[\frac{\delta_{\eta\lambda}\delta_{\gamma\delta}/2-2\frac{\kappa-1}{\kappa}\overline{n_\eta n_\lambda n_\gamma n_\delta}}{\kappa(\kappa+i\beta)}\right],\\\nonumber
\zeta_{\eta\lambda\gamma\delta}&=&\frac{j_0}{2}\int\limits_1^\infty\frac{(\kappa-1)d\kappa}{\kappa^2(\kappa+i\beta)}\\
\nonumber
&&\times
\frac{\partial}{\partial \kappa}\left[\frac{\delta_{\eta\lambda}\delta_{\gamma\delta}/2-2\frac{\kappa-1}{\kappa}\overline{n_\eta n_\lambda n_\gamma n_\delta}}{\kappa(\kappa+2i\beta)}\right]
\end{eqnarray}
are two auxiliary tensors, where the bar symbols stand for the averaging over the angle of the unity vector $\mathbf{n}$.
In Eq.~\eqref{EqTensors},
\begin{eqnarray}
j_0=\frac{(2e)^4}{2\pi \hbar^2 m}
\frac{T\beta^3}{\mu\omega^3}
E_1^2E_2,
\end{eqnarray}
and
\begin{eqnarray}
\label{Eqq13}
\overline{n_x n_x n_y n_y}
=\overline{n_y n_y n_x n_x }
=\overline{n_x n_y n_y n_x}
=\overline{n_y n_x n_x n_y}\\
\nonumber
=\overline{n_x n_y n_x n_y}=\overline{n_y n_x n_y n_x}=\frac{1}{8};~~~
\overline{n_x^4}=\overline{n_y^4}=\frac{3}{8},
\end{eqnarray}
whereas the other components (containing single $x$ or $y$ index such as $\overline{n_y n_x n_x n_x }$) vanish.
Expressions~\eqref{Eqq14}-\eqref{Eqq13} describe the general case of CPGE at any polarization and represent the main result of this paper.

\begin{figure}[tbp]
\includegraphics[width=0.49\textwidth]{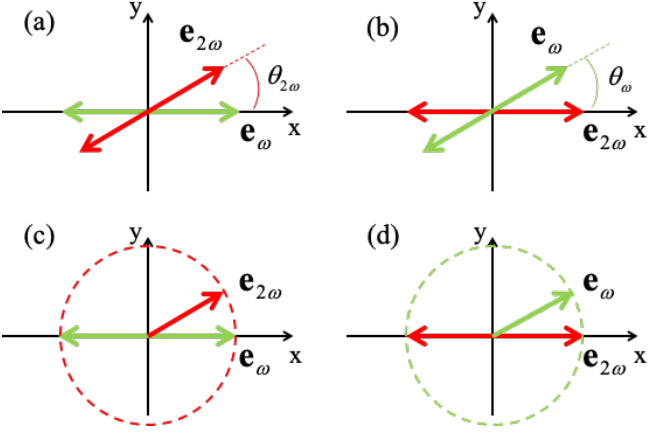}
\caption{Geometry of incident fields: Cases (a) and (b) correspond to linear polarization of both the fields, whereas (c) and (d) correspond to the cases of circular polarization of one of the fields (see text for details).}
\label{Fig1}
\end{figure}

\textit{Linear and circular polarizations.} Let us consider the most interesting cases from experimental point of view, presented in Fig.~\ref{Fig1}.
%
%
%
%
%
Choosing $\textbf{e}_{\omega}=(1,0)$ and $\textbf{e}_{2\omega}=(\cos\theta_{2\omega},\sin\theta_{2\omega})$ which corresponds to the case $(a)$ in Fig.~\ref{Fig1}, we find from Eqs.~\eqref{Eqq14}-\eqref{Eqq13},
\begin{eqnarray}\label{Eqq15}
j_x&=&2\cos\theta_{2\omega}\textmd{Re}\,(\chi_{xxxx}+\zeta_{xxxx}),
\\
j_y&=&2\sin\theta_{2\omega}\textmd{Re}\,(\chi_{yxxy}+\zeta_{yyxx}).
\end{eqnarray}

Instead, taking $\textbf{e}_{\omega}=(\cos\theta_{\omega},\sin\theta_{\omega})$ and $\textbf{e}_{2\omega}=(1,0)$ we find
for the case $(b)$,
\begin{eqnarray}\label{Eqq16}
j_x&=&2\cos^2\theta_{\omega}\textmd{Re}\,(\chi_{xxxx}+\zeta_{xxxx})\\
\nonumber
&&~~~~+2\sin^2\theta_{\omega}\textmd{Re}\,(\chi_{xyyx}+\zeta_{xxyy}),
\\
j_y&=&\sin (2\theta_\omega)\textmd{Re}\,(\chi_{yxyx}+\chi_{yyxx}+\zeta_{yxxy}+\zeta_{yxyx}).~~~
\end{eqnarray}
%
%


Furthermore, taking $\textbf{e}_{\omega}=(1,0)$ and $\textbf{e}_{2\omega}=(1,i\sigma)/\sqrt{2}$, where $\sigma=\pm 1$ indicates left/right circular polarization, we find for the case $(c)$ in Fig.~\ref{Fig1},
\begin{eqnarray}\label{Eqq17}
j_x&=&\sqrt{2}\textmd{Re}\,(\chi_{xxxx}+\zeta_{xxxx}),\\\nonumber
j_y&=&\sqrt{2}\sigma\textmd{Im}\,(\zeta_{yyxx}-\chi_{yxxy}).
\end{eqnarray}

Taking $\textbf{e}_{\omega}=(1,i\sigma)/\sqrt{2}$ and $\textbf{e}_{2\omega}=(1,0)$ for the case $(d)$ we have
\begin{eqnarray}\label{Eqq18}
j_x&=&\textmd{Re}\,(\chi_{xxxx}+\zeta_{xxxx}-\chi_{xyyx}-\zeta_{xxyy}),\\
\nonumber
j_y&=&\sigma\textmd{Im}\,(\chi_{yxyx}+\chi_{yyxx}-\zeta_{yxxy}-\zeta_{yxyx}).
\end{eqnarray}
%

%

\begin{figure}[tbp]
\includegraphics[width=0.49\textwidth]{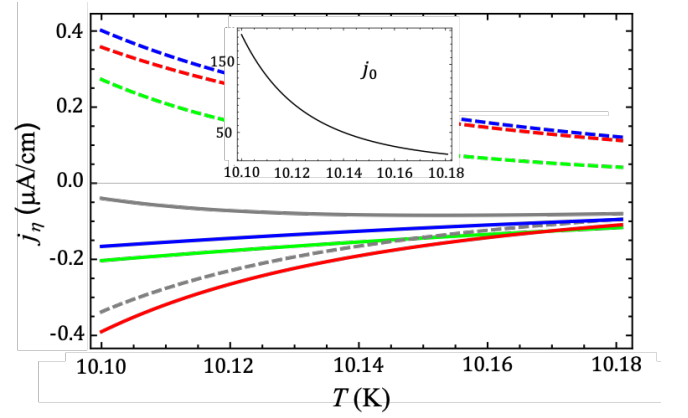}
\caption{Components of electric current densities, $j_x$ (solid) and $j_y$ (dashed curves), as functions of the temperature.
The green, gray, blue and red curves represent  (a), (b), (c), and (d) cases in Fig.~\ref{Fig1},  respectively.
Inset shows the dependence of $j_0$ on temperature. 
    We used $T_c=10$~K, $n=1.5\times 10^{11}$~cm$^{-2}$, $\tau=0.1$~ps, $m=0.5~m_0$, where $m_0$ is free electron mass.
    We fixed $\omega = 2 \times 10^{11} s^{-1}$,  $\theta_\omega=\theta_{2\omega}=\pi/6$,  $E_1=2$~V/cm, and $E_2=0.25$~V/cm.}
    \label{Fig3}
\end{figure}
\begin{figure*}
    \centering
    \includegraphics[width=0.8\textwidth]{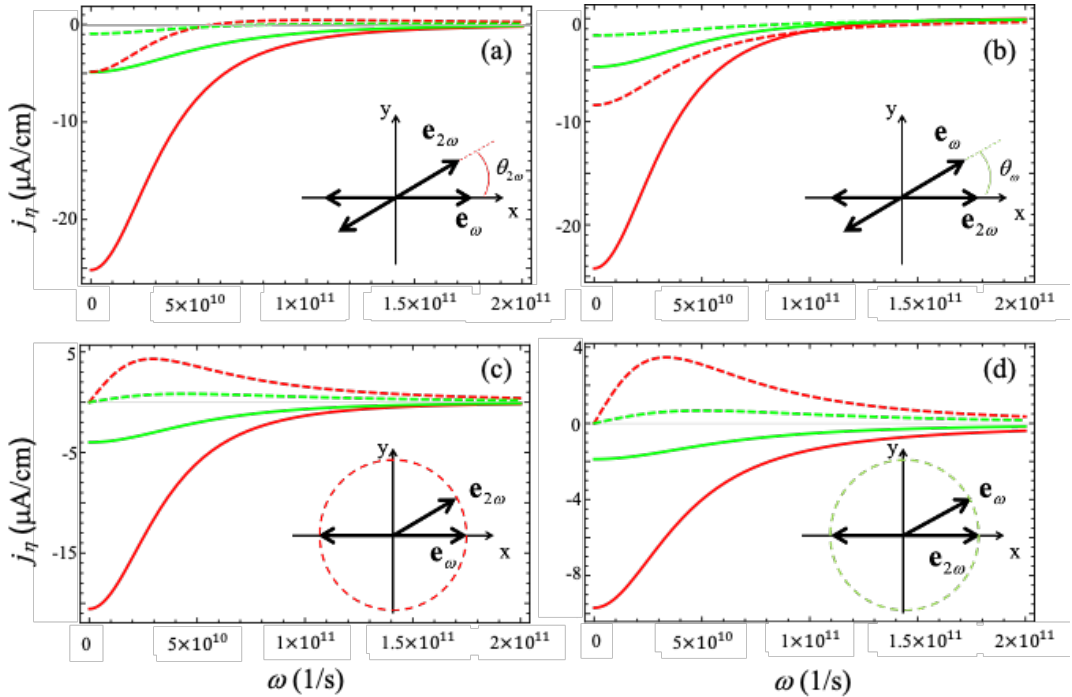}
    \caption{Spectra of electric current density for different incident field geometries.
    The solid curves show  $x$-projection of the current $j_x$ as  functions of frequency, while the dashed curves show $j_y$ as functions of the same frequencies at two temperatures: $10.10$~K (red) and $10.15$~K (green) that are above the critical temperature $T_c=10$~K.
    Other parameters were taken the same as in Fig.~\ref{Fig3}.}
    \label{Fig2}
\end{figure*}

\textit{Results and discussion.}
Figure~\ref{Fig3} shows the temperature dependence of the electric current density.
The components $j_x$ and $j_y$ corresponding to different cases in Fig.~\ref{Fig1} exhibit a decay once the temperature increases as compared with the critical temperature (compare also with Fig.~\ref{Fig2}).
It can be explained by the enhancement of the influence of SFs once we approach the critical temperature.
In general, the temperature dependence of the current is mainly (although not fully) inherited from the factor $j_0$, that goes as $T^{-2}$ at large temperatures, $T\gg T_c$, whereas it has a singularity $(T-T_c)^{-3}$ at temperatures approaching $T_c$ (see inset in Fig.~\ref{Fig3}).
This behavior is more singular than the one of the conventional paraconductivity, where the singularity is $(T-T_c)^{-1}$~\cite{LarkinVarlamov2005}.

Figure~\ref{Fig2} shows the spectra of electric current density corresponding to different geometries of incident fields, shown in Fig.~\ref{Fig1}.
In case of linearly polarized light [cases (a) and (b)], the x and y components of current density show a similar behavior.
The magnitude of current density decreases with the increase of frequency.
But this dependence is not monotonous in the case (a).
Indeed, at some frequency, the current density crosses zero and changes its sign, thus it starts to increase. Later it crosses a maximum and then decreases again.
Both the components of the current density saturate at high frequencies, independent of temperature.

In the case of circularly polarized light [panels (c) and (d) in Fig.~\ref{Fig2}], the x and y components of electric current density show different behavior.
While the x component behaves similar to that of linearly polarised light, the y components behave differently and reveal a non-monotonous behavior.
They also grow and then decay after overcoming a peak value.
With the decrease of $\epsilon$, the magnitudes of the currents grow.
We can also see that once the ambient temperature approaches $T_c$, the dashed curves ($j_y$) in panels (c) and (d) get narrower, and the peak frequency experiences a redshift.

\begin{figure*}[tbp]
\includegraphics[width=0.9\textwidth]{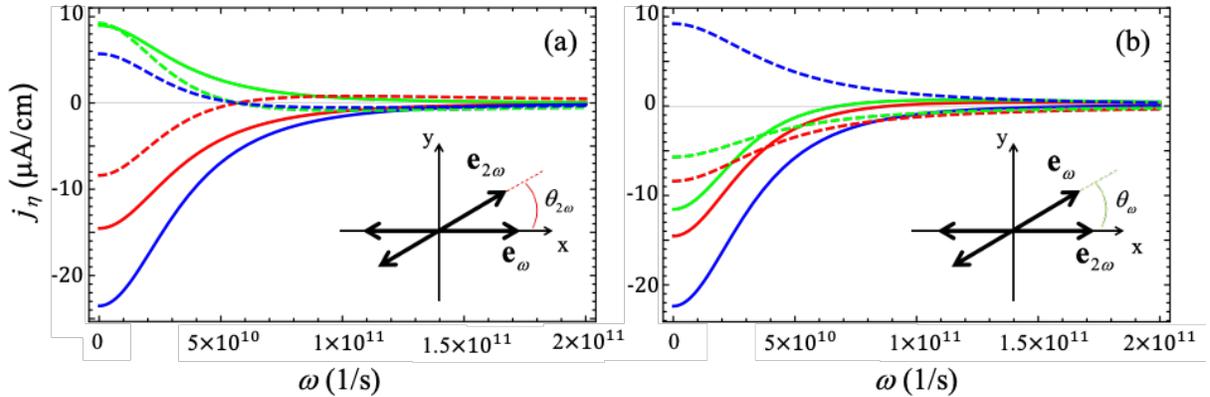}
\caption{Spectra of electric current densities corresponding to the cases (a) and (b) of Fig.~\ref{Fig1} for three different values of polarization angles $\theta_{2\omega}$ (a) and $\theta_{\omega}$ (b): $\pi/3$ (red), $7\pi/5$ (green), $9\pi/5$ (blue).
We used $T=10.1$~K.
Other parameters are taken the same as in Fig.~\ref{Fig3}.}
\label{Fig4}
\end{figure*}

Figure~\ref{Fig4} demonstrates the dependence of current densities on the polarization angles of the light fields.
As it follows from Eqs.~\eqref{Eqq15}-\eqref{Eqq18}, the $j_x$ and $j_y$ components of the current depend on the angle only in the cases when both the incident fields have linear polarization.
It corresponds to the cases (a) and (b) in Fig.~\ref{Fig1}.
As $\theta_{\omega}$ or $\theta_{2\omega}$ varies from 0 to $2\pi$,
the magnitudes of current densities change their magnitudes and even sign.
We want to note, that for the cases (c) and (d) in Fig.~\ref{Fig1}, there is an extra factor $\sigma$, which reflects the chirality of the field but there is no dependence on the angle.

In this article, we have considered one particular type of Superconducting fluctuations: the Aslamazov-Larkin corrections.
There also take place other contributions: the Maki-Thompson~\cite{Maki,Thomson} and the ``density of states''~\cite{ALDOS} corrections.
However, the Boltzmann equations approach cannot be used for their description, and a quantum approach is required.


\textit{In conclusion,} we have studied the coherent photogalvanic effect in a two-dimensional  superconductor in the fluctuating regime.
We have shown the emergence  of a stationary electric current of Cooper pairs when the sample is exposed to two coherent electromagnetic fields of light with certain frequencies and different polarizations.
We have derived the general formula for the electric current density and investigated in detail the particular cases of linear and circular polarizations of the external light fields.
We have shown, that the current might experience a non-monotonous dependence on frequency and it is very sensitive to the proximity of the temperature to the critical temperature of superconducting transition. In particular, the peak in the spectrum of the current becomes narrower, its frequency experiences a redshift, and the intensity of the peak grows once the temperature approaches $T_c$.
These results capture the effects arising due to the interplay of the physics of superconducting (Cooper pair density) fluctuations and the polarizations of incident light fields.

\textit{Acknowledgements.} We thank M. Entin for fruitful discussions and critical reading of the text. 
We acknowledge the support by the Institute for Basic Science in Korea (Project No.~IBS-R024-D1) and the Russian Foundation for Basic Research (Project No.~18-29-20033).

%
%
%
%
%
%



%
%

%
%
%
%
%
%


\end{document}